\documentclass[11pt,a4paper]{article}

\usepackage{amsmath,amssymb}
\usepackage{epsfig,graphicx}
\usepackage{subfigure}
\usepackage{graphicx}
\usepackage{rotating}
\usepackage{cancel}
\usepackage{bm}
\usepackage{color}
\usepackage{comment}
\usepackage{cite}

\renewcommand\[{\left[}

\newcommand{\exclude}[1]{}

\def\beq{\begin{equation}}
\def\eeq{\end{equation}}

\begin{document}
\numberwithin{equation}{section}
\title{{\normalsize  \mbox{}\hfill~}\\
\vspace{2.5cm} 
\Large{\textbf{Directional Resolution of Dish Antenna Experiments \\[0.1cm]
to Search for WISPy Dark Matter} \vspace{0.5cm}}}

\author{Joerg Jaeckel and Stefan Knirck\\[2ex]
\small{\em Institut f\"ur theoretische Physik, Universit\"at Heidelberg,}\\
\small{\em Philosophenweg 16, 69120 Heidelberg, Germany}\\[0.5ex]
}

\date{}
\maketitle

\begin{abstract}
\noindent
Dark matter consisting of very light and very weakly interacting particles such as axions, axion-like particles and hidden photons
could be detected using reflective surfaces. On such reflectors some of the dark matter particles are converted into photons and, given a suitable geometry, concentrated on the detector. This technique offers sensitivity to the direction of the velocity of the dark matter particles. In this note we investigate how far spherical mirrors can concentrate the generated photons and what this implies for the resolution in directional detection as well as the sensitivity of discovery experiments not aiming for directional resolution. Finally we discuss an improved setup using a combination of a reflecting plane with focussing optics.
\end{abstract}

\newpage

\section{Introduction}
Our Universe contains five times more dark matter than ordinary matter, and dark matter contributes 27\% to the total energy content.
Yet its nature still remains unknown. Indeed there is still room for very different hypothesis for what it could be.
This is exemplified by the fact that it could be WIMPs~\cite{Bertone:2004pz}, weakly interacting massive particles, with masses of hundreds of GeV or even TeV, or it could be WISPs~\cite{Arias:2012az}, (very) weakly interacting slim particles, with masses in the eV range or even many orders of magnitude lighter than that (see, e.g.~\cite{Jaeckel:2010ni} for a review).
Consequently to detect dark matter we must look in very different direction, using a wide variety of techniques.

In this note we consider a recently proposed technique~\cite{Horns:2012jf} for detecting WISPy dark matter particles, in particular axion(-like particle)s and hidden photons using suitably shaped reflecting surfaces. A first experiment has already been performed~\cite{Suzuki:2015sza} and a second more sensitive experiment is already under way~\cite{Dobrich:2014kda}. The main mechanism (details below) is that reflecting surfaces convert these particles into ordinary photons emitted essentially perpendicular to the surface. Using a spherically shaped reflector the produced photons are concentrated in the centre of the sphere, where they can then be detected.

Importantly the emission of photons is not always exactly perpendicular to the surface. This only holds when the dark matter particles that are converted are at rest with respect to the surface. Dark matter particles moving with a velocity of order $v\sim 10^{-3}$ parallel to the surface are emitted at an angle $\sim v$ with respect to the normal direction. This can be used for directional detection~\cite{Jaeckel:2013sqa} but it also has important implications for the design of the experiment~\cite{Jaeckel:2013sqa}. In particular, since dark matter particles have a velocity distribution with a width of the order of $\sim 10^{-3}$ the detector needs to have a sufficient size to capture a significant part of the produced photons.

The imaging properties of a spherical mirror are, however, not perfect with respect to making an image of the velocity distribution. The main aim of this note is to quantify the deviations from perfect imaging and determine resulting limitations on the resolution.

In particular, in~\cite{Jaeckel:2013sqa} only the case of a nearly flat mirror of a diameter $D$ much smaller than the curvature radius $R$, $D\ll 2R$ was considered. 
In real experiments like~\cite{Suzuki:2015sza} and the FUNK experiment~\cite{Dobrich:2014kda} this is only a first approximation. In the following we will take the FUNK experiment as our benchmark example. The FUNK mirror has a size of $13\,{\rm m}^2$ and a curvature radius of $R=3.4\,{\rm m}$. For a spherical mirror\footnote{Indeed the mirror of the FUNK experiment actually has a more quadratic shape such that the maximum diameter is actually larger.} of this size one finds
\begin{equation}
\frac{D}{2R}\sim 0.6
\end{equation}
which suggests that sizeable deviations from a mirror with $D\ll 2R$ are possible. 
This is what we want to quantify.

The note (which is in large parts based on~\cite{Knirck}) is structured as follows. In the following Sect.~\ref{conversion} we recall the conversion of WISPy dark matter moving with a velocity $v$ into photons. Then in Sect.~\ref{spherical} we discuss directional detection with a spherical mirror beyond the limit where the mirror is small compared to the radius of curvature. We also comment on higher order corrections in $v$. However, we will remain in the limit of geometric optics, i.e. the wavelength must be small compared to the mirror size.
In Sect.~\ref{parabolic} we investigate a different setup consisting of a plane and a parabolic mirror. We summarize and conclude in Sect.~\ref{conclusions}.

\section{Photon emission from the conversion of moving dark matter particles}\label{conversion}
In the following we will assume that the wavelength of the emitted photons is small enough such that diffraction effects can be neglected. This requires that the mass $m$ of the dark matter particles is sufficiently big, $1/m\ll D$, where $D$ is the diameter of the mirror.

Moreover for simplicity we will focus on the case of hidden photon dark matter~\cite{Horns:2012jf,Jaeckel:2013sqa}. 
We note, however, that axion-like particles can be detected in a very similar way provided a magnetic field is present at the reflector surface\footnote{Indeed the resulting equations are essentially the same, with the replacement, $\chi\rightarrow \chi_{\rm eff}=\frac{g_{a\gamma\gamma}|\mathbf{B}|}{m_{a}}$.}.

Let us briefly recall the basics of hidden photon dark matter~\cite{Nelson:2011sf,Arias:2012az} and its conversion on a reflecting surface~\cite{Horns:2012jf,Jaeckel:2013sqa}.  Hidden photons can kinetically mix~\cite{Holdom:1985ag} (see, e.g.~\cite{Jaeckel:2013ija} for a review) with the ordinary photon,
\begin{equation}
{\mathcal{L}}=-\frac{1}{4}F^{\mu\nu}F_{\mu\nu}-\frac{1}{4}X^{\mu\nu}X_{\mu\nu}-\frac{\chi}{2}F^{\mu\nu}X_{\mu\nu}+\frac{m^{2}_{X}}{1+\chi^2}X^{\mu}X_{\mu}+j^{\mu}A_{\nu},
\end{equation}
where $F^{\mu\nu}$ and $X^{\mu\nu}$ are the ordinary and hidden photon field strength linked to the gauge fields $a^{\mu}$ and $X^{\mu}$, respectively. $m^{2}_{X}$ is the mass of the hidden photons and $j^{\mu}$ denotes the coupling to matter as in our case the mirror.
Typical values for the kinetic mixing parameters in extensions of the Standard Model range from $10^{-12}$ to $10^{-3}$~\cite{Holdom:1985ag,Dienes:1996zr}.

The average density of hidden photons moving with different momenta ${\mathbf{k}}$ is given by,
\begin{equation}
\rho_{\rm HP}=\frac{1}{2} m^{2}_{X}\int \frac{d^{3} {\mathbf{k}}}{(2\pi)^3}\langle |{\mathbf{X}}_{\rm DM}({\mathbf{k}})|^2\rangle.
\end{equation}
If hidden photons are all of the dark matter this fixes
\begin{equation}
\rho_{\rm HP}=\rho_{\rm CDM}.
\end{equation}

Due to the kinetic mixing term the massive (mostly hidden photon) eigenstate couples also to the photon with a coupling proportional to $\chi$.
Let us consider a single momentum mode of our dark matter particles corresponding to a plane wave of hidden photons with momentum ${\mathbf k}_{0}$. We then have for the effective electric field components,
\begin{equation}
\label{hpwave}
\left(\begin{array}{c}
{\mathbf E}
\\
{\mathbf E}_{\rm hid}
\end{array}\right)
={\mathbf E}_{\rm DM}\left(\begin{array}{c}
-1
\\
1/\chi
\end{array}
\right)\exp(-i(\omega t -{\mathbf k}_{0}{\mathbf x})).
\end{equation}
Here the field ${\mathbf E}$ denotes the components coupling to the ordinary charges, and ${\mathbf E}_{\rm hid}$ those that couple only to hidden charges, which we assume to be absent in our setup. 
The size of the electric field is related to the hidden photon amplitude via
\begin{equation}
{\mathbf{E}}_{\rm DM}=\chi m_{X}{\mathbf{X}}_{\rm DM}.
\end{equation}

\subsection*{Emission from a plane mirror}
Let us consider a wave as in Eq.~\eqref{hpwave} impinging on an ideal conducting plane surface (or a mirror). Then the components of the ordinary electric field
in the directions parallel to the plane are eliminated by the movement of the electrons. This causes the emission of an ordinary electromagnetic
wave,
\begin{equation}
\left(\begin{array}{c}
{\mathbf E}
\\
{\mathbf E}_{\rm hid}
\end{array}\right)_{\rm emitted}={\mathbf E}_{{\rm DM},\parallel}\left(
\begin{array}{c}
1
\\
\chi
\end{array}\right)\exp(-i(\omega t-{\mathbf{k}}_{1}{\mathbf{x}})),
\end{equation}
such that the boundary condition for the ordinary electromagnetic field parallel to the plane,
\begin{equation}
\label{boundary}
0={\mathbf E}_{\rm tot, \parallel}|_{\rm surface},
\end{equation}
is fulfilled. 

For our purposes the most important question is now the relation between the velocity of the incoming dark matter particle and that of the emitted electromagnetic wave. This can be obtained as follows.
To fulfil the relation~\eqref{boundary} everywhere on the surface requires for the momentum ${\mathbf{k}}_{1}$ of the outgoing photon wave
\begin{equation}
{\mathbf{k}}_{\parallel,1}={\mathbf{k}}_{\parallel,0}.
\end{equation}
This result can also be easily understood from momentum conservation. Since the system has translational invariance in the 
plane, momentum in these directions must be conserved.

The remaining component of ${\mathbf{k}}_{1}$ can be determined by energy conservation,
\begin{equation}
|{\mathbf{k}}_{0}|^2+m^{2}_{X}=\omega^2=|{\mathbf{k}}_{1}|^2.
\end{equation}

With this we can obtain the outgoing wave vector,
\begin{equation}
{\mathbf{k}}_{1}=\sqrt{m^{2}_{X}+|{\mathbf{k}}_{\perp,0}|^2}\,\,{\mathbf{n}}+{\mathbf{k}}_{\parallel,0},
\end{equation}
where ${\mathbf n}$ is the normal vector of the reflecting surface.

\begin{figure}[t]
\centering
   \includegraphics[width=3.5cm]{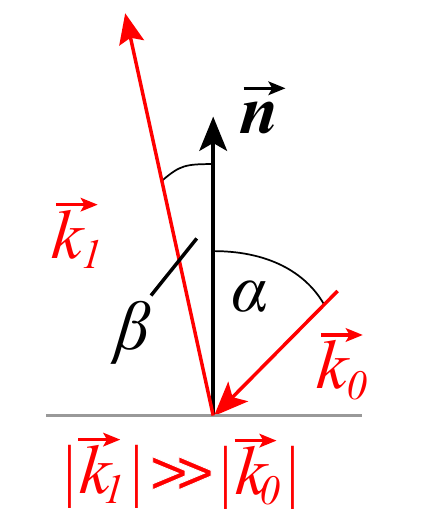} 
   \caption{Definition of the angles for the incoming dark matter particle and the outgoing photon wave (figure from~\cite{Knirck}).}
   \label{geometry}
\end{figure}

Using these equations we can derive a simple law for the relation between the angles of the incoming hidden photon and the outgoing proper photon wave (cf. also Fig.~\ref{geometry}),
\begin{equation}
\label{inout}
\sin(\beta)=\sin(\alpha)\frac{v}{\sqrt{1+v^2}},
\end{equation}
where
\begin{equation}
v=\frac{|{\mathbf{k}}_{0}|}{m_{X}}.
\end{equation}

\bigskip 

For completeness we note that the emitted photon power $dP$ per area $dA$ is given by
\begin{equation}
\frac{dP}{dA}=\langle \cos^{2}(\alpha)\rangle \chi^2 \rho_{\rm CDM},
\end{equation}
where $\alpha$ denotes the angle between the hidden photon polarization and the surface element and the average is taken over the
dark matter distribution\footnote{With a suitable setup, measuring with mirrors at different angles, one could even think of measuring the polarziation
distribution of the hidden photons.}.

\section{Spherical mirror}\label{spherical}
For $v=0$ Eq.~\eqref{inout} yields perpendicular emission from the surface. A spherical mirror will therefore
concentrate the whole signal in the centre of the sphere~\cite{Horns:2012jf}.
However, in our galaxy dark matter is not entirely at rest, but one expects that it moves with a velocity of the order of the virial velocity,
\begin{equation}
v\sim 10^{-3}.
\end{equation}
In this case the relation Eq.~\eqref{inout} is not trivial anymore and the rays will be spread over some area of size $\sim v R$, where $R$ is the curvature radius. As discussed in~\cite{Jaeckel:2013sqa} this can be used for directional detection if one uses a finite area detector with spatial resolution.

\subsection*{Emission from a single surface element}
\begin{figure}[t]
\centering
   \includegraphics[width=10cm]{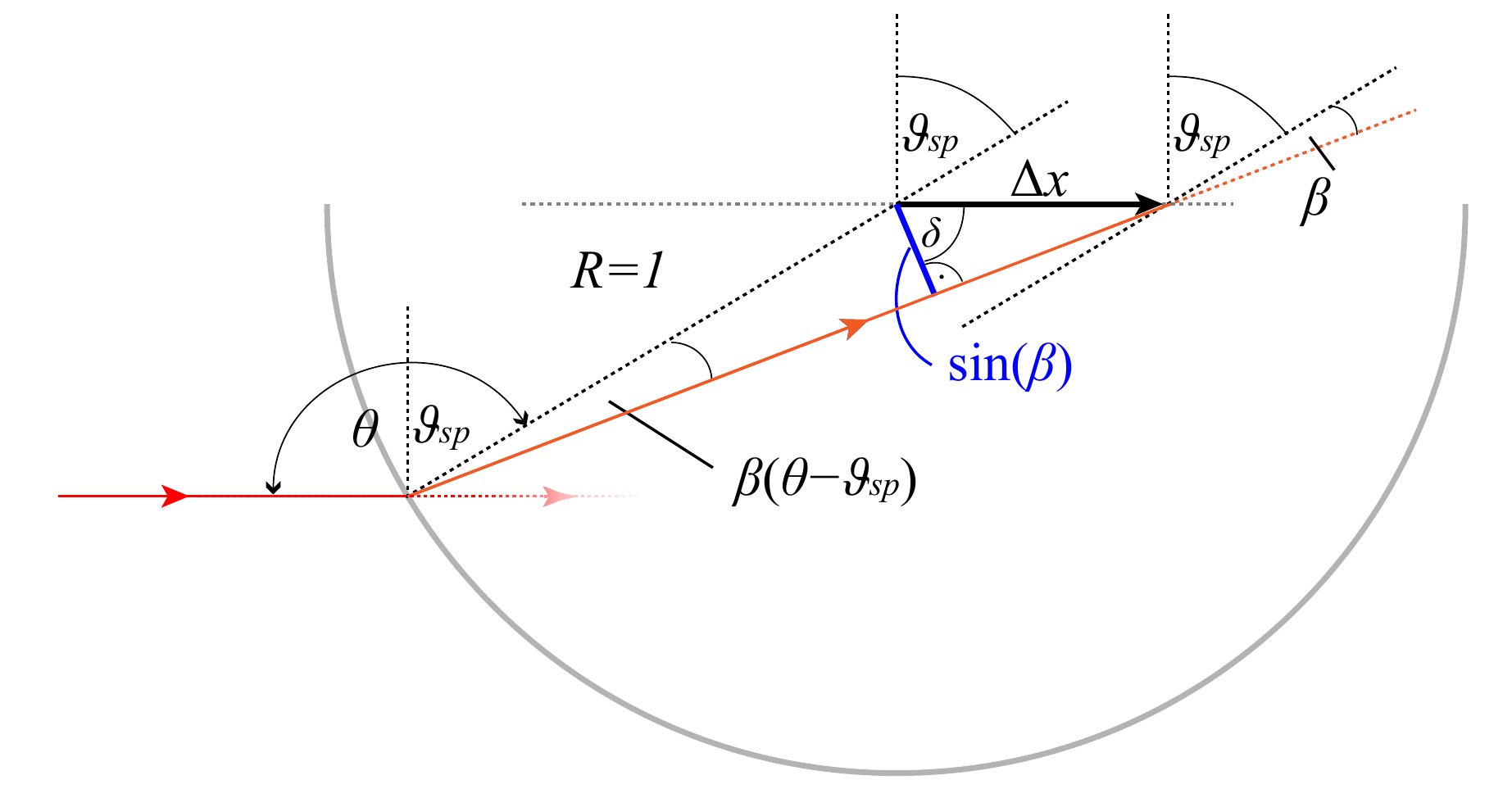} 
   \caption{Definition of angles and geometrical consideration for the displacement of the photon impact on the detector in two dimensions. $\theta$ is the angle between the incoming hidden photon wave and the main axis of the mirror (the case shown is $\theta=90^{\circ}$). $\vartheta_{sp}$ characterizes the location on the sphere where the hidden photon is converted.  $\Delta x$ is the displacement of the impact on the detector from the centre of the sphere. Figure from~\cite{Knirck}.}
   \label{2dsphere}
\end{figure}

Let us start with a simple consideration in two dimensions as shown in Fig.~\ref{2dsphere}. 
The main quantity of interest is the displacement $\Delta x$ of the detected photon from the centre of the sphere.
From the geometry shown in Fig.~\ref{2dsphere} one reads off
\begin{equation}
\Delta x=R\frac{\sin(\beta)}{\cos(\vartheta_{sp}+\beta)}=R\frac{\sin(\beta)}{\cos(\vartheta_{sp})\cos(\beta)-\sin(\vartheta_{sp})\sin(\beta)}.
\end{equation}
One can now insert Eq.~\eqref{inout} to obtain a relation between the direction and velocity of the incoming dark matter particle and the displacement on the detector.

For a mirror with small diameter $\vartheta_{sp}\ll 1$ for all points on the mirror, neglecting $\vartheta_{sp}$ and using the first order approximation
in $v$ one quickly obtains
\begin{equation}
\Delta x \approx v\sin(\theta) R,
\end{equation}
which agrees with the result of~\cite{Jaeckel:2013sqa}.

Let us now go two steps further in the approximation. We keep $\vartheta_{sp}$ completely arbitrary but expand to second order in $v$.
We then have
\begin{eqnarray}
\label{approx}
\Delta x &=&R\bigg[+\sin(\theta) v
\\\nonumber
&&
\quad\,\,\,-\tan(\vartheta_{sp})\cos(\theta)v
\\\nonumber
&& \quad\,\,\,+\left[\sin(\vartheta_{sp})\left[\tan^{2}(\vartheta_{sp})\cos^2(\theta)+\sin^{2}(\theta)\right]-\cos(\vartheta_{sp})\tan^{2}(\vartheta_{sp})\sin(2\theta)\right]v^2
+\ldots  \bigg].
\end{eqnarray}
The first line is the leading order result. 
Importantly, for $\vartheta_{sp}\neq 0$ there is a correction, given in the second line, that appears already at leading order in $v$. 
The third line then gives the corrections of order $v^2$.

Let us now discuss the relevance of the various terms. The terms in the first two lines are both of the same order in $v$. Barring extreme angles the size of the correction given in the second line is determined by $\tan(\vartheta_{sp})$. Let us estimate this for a mirror of the size of the one used in the FUNK experiment~\cite{Dobrich:2014kda}. With an are of 13~m$^2$ and a curvature radius of $3.4$~m, and using the simplifying assumption of a simple spherical cap geometry, one finds a maximal $\vartheta_{sp, max}=0.6$ and a corresponding 
\begin{equation}
\tan(\vartheta_{sp, max})\sim 0.7\qquad {\rm FUNK\,\,geometry}.
\end{equation}
Therefore the correction in the second line can not be neglected. 

Let us now turn to the corrections of order $\sim v^2$. Compared to the leading term and the corrections in the second line those terms are suppressed by a factor of $v\sim 10^{-3}$. For small angles $\vartheta_{sp}$ they are therefore completely negligible.
For large $\vartheta_{sp}$ such that $\tan(\vartheta_{sp})>1$ they are at worst enhanced by another factor of $\tan(\vartheta_{sp})$.
Unless\footnote{The region where $\tan(\vartheta_{sp})\gtrsim 1/v$.} $\vartheta_{sp}=\pi/2-{\mathcal{O}}(v)$, i.e. the mirror is essentially a complete half sphere, these corrections are very small over the entire area of the mirror and can be neglected. Even if the mirror is a full half sphere, the corrections are still small, because the area which feature a sufficiently high $\vartheta_{sp}$ is only the very rim of the half sphere and quite small (we have also checked this numerically~\cite{Knirck}).
\bigskip

These results straightforwardly generalize to a full three dimensional situation. 
Using the rotational symmetry of the problem about the z-axis we can, without loss of generality, assume that the incident particles have non-vanishing velocity only in the y- and z-direction.
The impact of the photons from the conversion then occurs at
\begin{equation}
\label{approx3d}
\Delta\mathbf{x}=\left(\begin{array}{c}
\Delta x
\\
\Delta y
\end{array}\right)
=R\left[\left(\begin{array}{c}
0
\\
1
\\
\end{array}
\right) \sin(\theta)v
-
\left(\begin{array}{c}
\cos(\phi_{sp})
\\
\sin(\phi_{sp})
\end{array}\right)v\tan(\vartheta_{sp})\cos(\theta)v+{\mathcal{O}}(v^2)\right],
\end{equation}
where $\phi_{sp}$ denotes the azimuthal angle of the emitting surface element.
The ${\mathcal{O}}(v^2)$ corrections are negligible.
\bigskip

In summary, for the expected dark matter velocities of the order of $v\sim 10^{-3}$ corrections of the order $\sim v^2$ are negligible for reasonable geometries. However, corrections from the non-vanishing angular extent of the mirror given by the second part on the right hand side of Eq.~\eqref{approx3d} can be quite significant for realistic mirror geometries such as used in the FUNK experiment.

\subsection*{Signal integration over the mirror area}
So far we have considered only the signal from a single point of the mirror at a location specified by $(\vartheta_{sp},\phi_{sp})$.
In order to obtain the signal of the full mirror we have to integrate over its whole area.

For simplicity we still consider dark matter incoming with a velocity $v$ at a fixed incident angle $\theta$. Moreover, we consider a spherical cap mirror
with opening angle $\vartheta_{sp,max}$.
The expected intensity distribution on the detector is then given by integrating over all the surface elements,
\begin{equation}
I(x,y)=\frac{1}{\mathcal{N}}\int^{\vartheta_{sp, max}}_{0}d\vartheta_{sp}\,\int^{2\pi}_{0}\sin(\vartheta_{sp})d\phi_{sp}\,\delta(x-\Delta x(\vartheta_{sp},\phi_{sp}))\delta(y-\Delta y(\vartheta_{sp},\phi_{sp}),
\end{equation}
where ${\mathcal{N}}$ denotes a suitable normalization.

Inserting Eq.~\eqref{approx3d} this can be evaluated to
\begin{equation}
\label{dist}
I(x,y)=\bigg\{\begin{array}{lcl}
\frac{1}{\mathcal{N}}\frac{|\cos(\theta) v|}{\left(x^2+\left[y-R\sin(\theta)v\right]^2+R^2\cos^{2}(\theta)v^2\right)^{3/2}} & {\rm for}& 
\vartheta_{0}\leq \vartheta_{sp, max}
\\
0 & {\rm for} & \vartheta_{0}> \vartheta_{sp, max}
\end{array}
\end{equation}
with
\begin{equation}
\vartheta_{0}=\left|\arctan\left(\frac{\sqrt{x^2+(y-R\sin(\theta)v)^2}}{R\cos(\theta)v}\right)\right|.
\end{equation}

\subsection*{Resolution}
Looking at the distribution Eq.~\eqref{dist} we find that the maximum intensity occurs at the location,
\begin{equation}
x=0\qquad y=R\sin(\theta)v,
\end{equation}
that is expected from the lowest order approximation in the opening angle of the mirror $\vartheta_{sp, max}\ll 1$.
However, for finite opening angles $\vartheta_{sp,max}$ we now have  a distribution over a finite area of the detector instead of a $\delta$-function like peak.
To quantify this area we can consider the detector region where the intensity is greater than half the maximal intensity.

Looking at Eq.~\eqref{dist} we find that the intensity only depends on the distance squared from the maximum of the distribution,
\begin{equation}
\Delta^2=x^2+\left[y-R\sin(\theta)v\right]^2.
\end{equation}
If the opening angle $\vartheta_{sp,max}$ is large $\vartheta_{sp,max}\sim \pi/2$ we only need to consider the first line of Eq.~\eqref{dist}.
Half intensity is reached, when
\begin{equation}
\Delta=(2^{2/3}-1)^{1/2}R\cos(\theta)v\approx 0.77 R\cos(\theta) v.
\end{equation}
Accordingly the spread is roughly of the same size as the displacement of the peak from the centre $\Delta y=R\sin(\theta)v$.

Indeed if we take $\Delta$ as the uncertainty with which we can measure the location of the peak this corresponds to an uncertainty in the velocity
parallel to the mirror plane $\Delta\mathbf{v}_{\parallel}$, relative to the total velocity $v$,
\begin{equation}
\frac{|\Delta\mathbf{v}_{\parallel}|}{v}=\frac{\Delta}{Rv}=0.77\cos(\theta).
\end{equation}
Velocity resolution is therefore strictly limited. Note, that the width depends on $\cos(\theta)$, i.e. the velocity component perpendicular to the surface.

On the positive side, the spread is also not much larger than a typical displacement. For a discovery experiment, that does not aim for precise directional resolution, one can use a mirror with large opening angle without the need to significantly increase the detector size (see below for details).

Let us now restrict the opening angle of the mirror $\vartheta_{sp,max}$ and discuss its effect on the resolution.
From Eq.~\eqref{dist} we can see that for $\vartheta_{0}=\vartheta_{sp,max}$ the intensity drops to zero. This occurs when
\begin{equation}
\Delta =R\tan(\vartheta_{sp,max})\cos(\theta)v.
\end{equation}
All photons are contained within this radius of the peak.

Using this the relative velocity uncertainty is therefore
\begin{equation}
\label{uncert}
\frac{|\Delta\mathbf{v}_{\parallel}|}{v}=\frac{\Delta}{Rv}=\tan(\vartheta_{sp,max})\cos(\theta).
\end{equation}
Significantly improved resolution can only be obtained when
\begin{equation}
\tan(\vartheta_{sp,max})\ll 0.77.
\end{equation}

\subsection*{Implications for discovery experiments}
Let us also look at the possible implications for discovery experiments that do not aim for directional resolution.
The effect of non-ideal imaging for those experiment is that the photons are spread over a wider area. For the necessarily
finite detectors this may lead to a loss of photons.

Our above considerations can be used to determine the conservative detector area that catches essentially all the emitted photons, taking into account the non-ideal imaging properties of a spherical mirror.
If the maximal dark matter velocity with respect to the setup is given by $v_{max}$, a detector of size
\begin{equation}
r_{det}=(1+\tan(\vartheta_{sp,max}))Rv_{max}
\end{equation}
is sufficient. For our example of the FUNK detector the non-ideal imaging suggests to increase the detector radius by about 70\%.

However, this is overly conservative since a significant fraction of the photons will actually arrive at a point closer to the centre of the detector than suggested by ideal imaging. Moreover, for more or less isotropic velocity distributions and detectors that cover a sizeable part of the velocity distribution, the typical dark matter particles impacting at the boundaries of the detector will have a relatively small velocity perpendicular to the mirror plane (since they have near maximal velocity in the parallel directions). As we can see from Eq.~\eqref{uncert} the spread depends on the perpendicular velocity 
($\sim v\cos(\theta)$)and is therefore smaller at those boundaries.

\begin{figure}[t]
\centering
   \includegraphics[width=7cm]{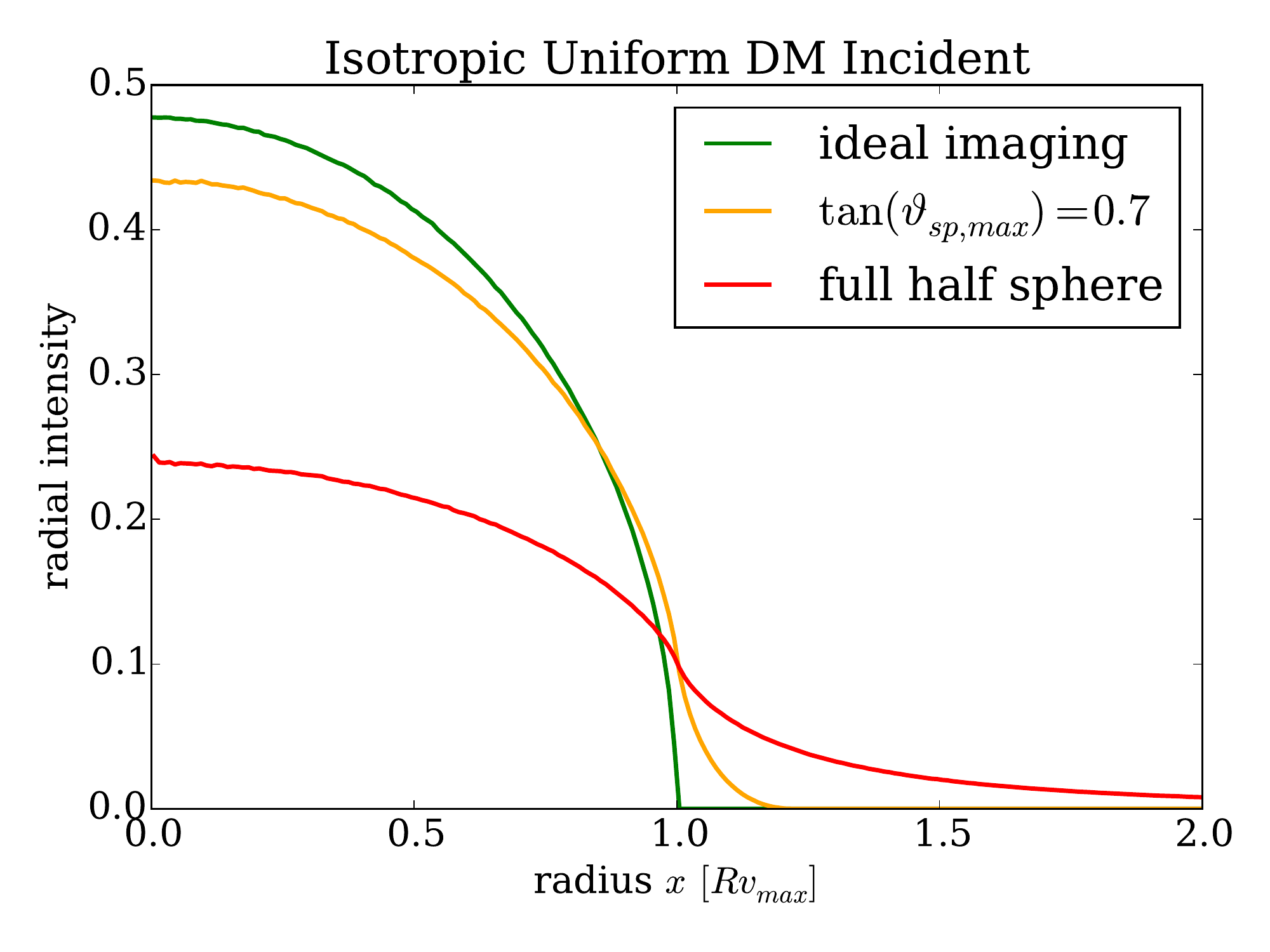} 
   \hspace*{0.5cm}
   \includegraphics[width=7cm]{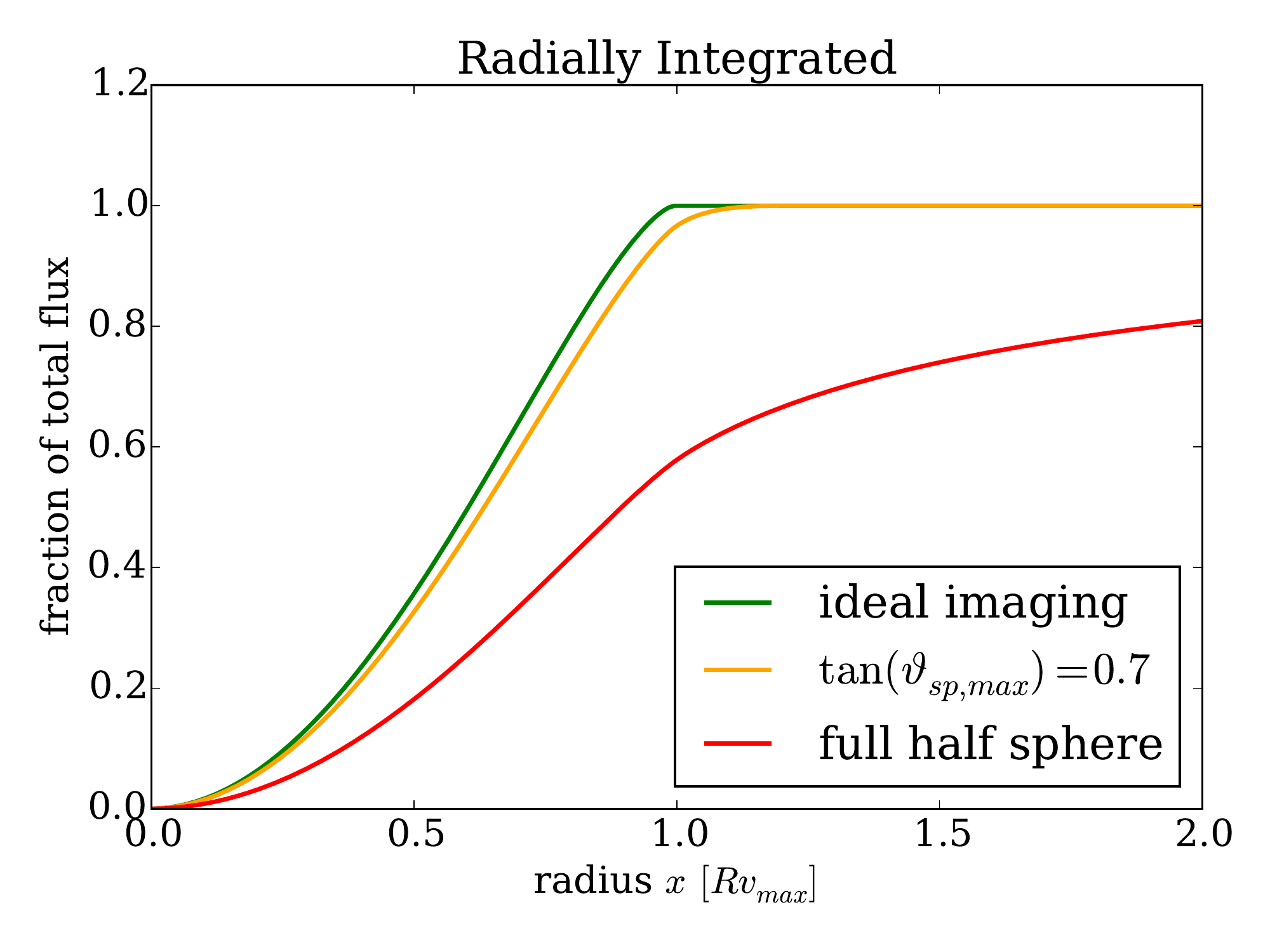} 
   \caption{{\bf Left Panel:} Radial intensity distribution for an example dark matter velocity distribution. The distributions are normalized such that the integral over the intensity is equal to~$1$ in each case. {\bf Right Panel:} Fraction of the total flux captured within a radius $x$ of the detector.}
   \label{distribution}
\end{figure}

As an explicit example we plot in the left panel of Fig.~\ref{distribution} the radial intensity distributions for an example dark matter velocity distribution,
\begin{equation}
\label{uniform}
f(\mathbf{v})=f_{0}\,\theta(v_{max}-|\mathbf{v}|).
\end{equation}
We show ideal imaging in green, a mirror with $\tan(\vartheta_{sp, max})=0.7$ in yellow and the full half sphere in red. 
The right panel shows the fraction of the total flux that is contained within a radius $x$.
We can see that, for the chosen velocity distribution, the difference in the fraction of the total flux for ideal imaging and a real mirror like the one employed in FUNK is not dramatic. For reasonable radii of the detector $\sim Rv_{\rm max}$, typically about 90\% of the ideal flux are captured, even when non-ideal imaging is included. Nevertheless, we note that this is specific for the given dark matter distribution.
For a proper determination of the sensitivity the non-ideal imaging will have to be taken into account and different dark matter distributions will have to be studied.

\begin{figure}[t]
\centering
   \includegraphics[width=7cm]{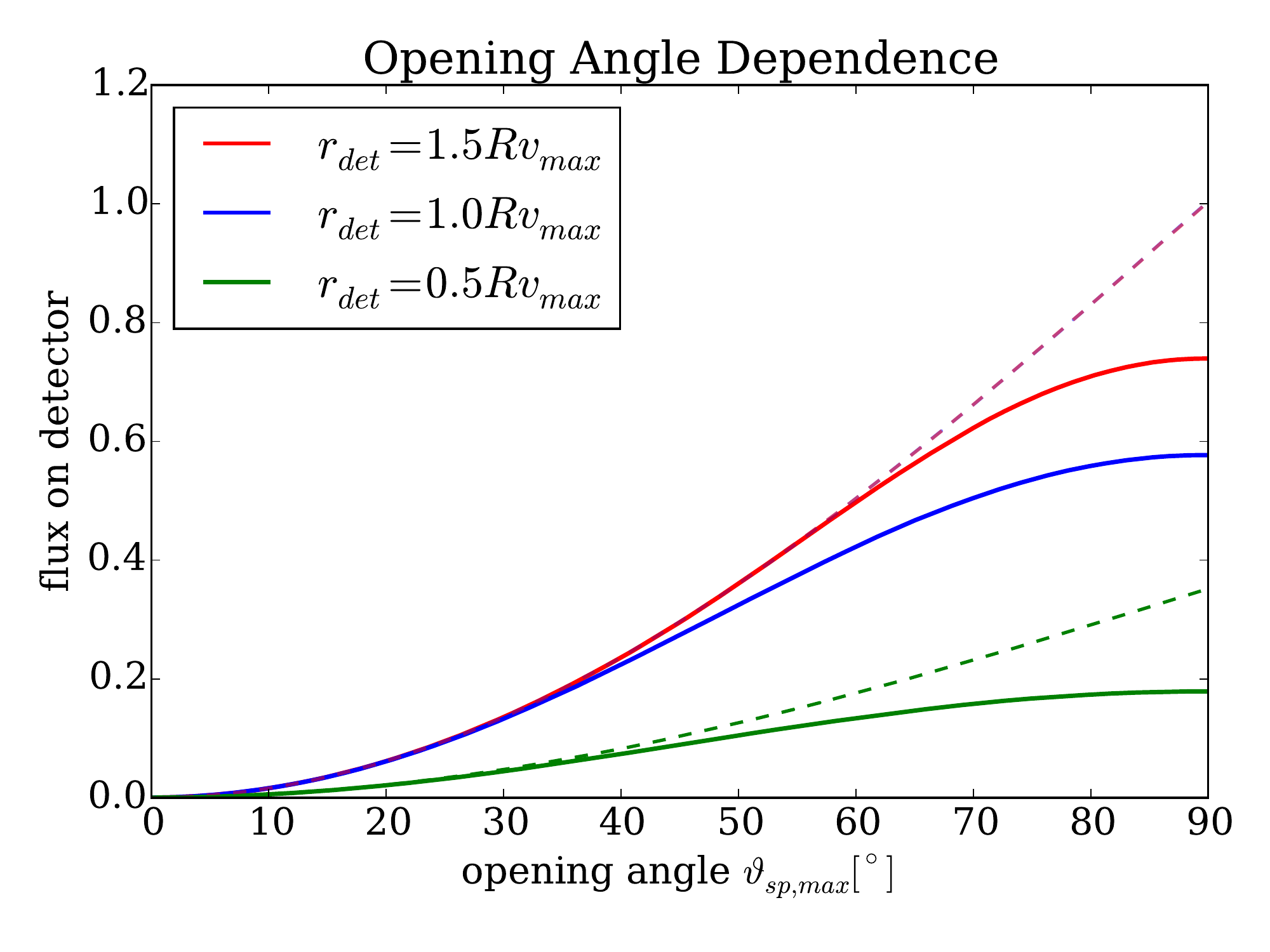} 
   \caption{Total flux as a function of the opening angle $\vartheta_{sp, max}$ for three different detector sizes. Dashed lines show the results assuming ideal imaging. The flux is normalized to the total number of photons produced by a full half sphere. We note that the red and blue dashed lines are on top of each other, since for ideal imaging no photons impact at a distance greater than $Rv_{max}$.}
   \label{totalflux}
\end{figure}

In Fig.~\ref{distribution} we show the fraction of all the photons produced by the mirror that impact on the detector.
However, for a discovery experiment what truly counts is the total flux impacting on the detector. 
As we increase the opening angle, keeping curvature radius and detector size fixed, the total area increases and additional photons will reach the detector. We show this in Fig.~\ref{totalflux}, where we plot the photon flux for our example distribution as a function of opening angle and for different detector sizes. 
If one does not aim for directional detection the flux is always maximal for maximal opening angle.

\begin{figure}[t]
\centering
   \includegraphics[width=7cm]{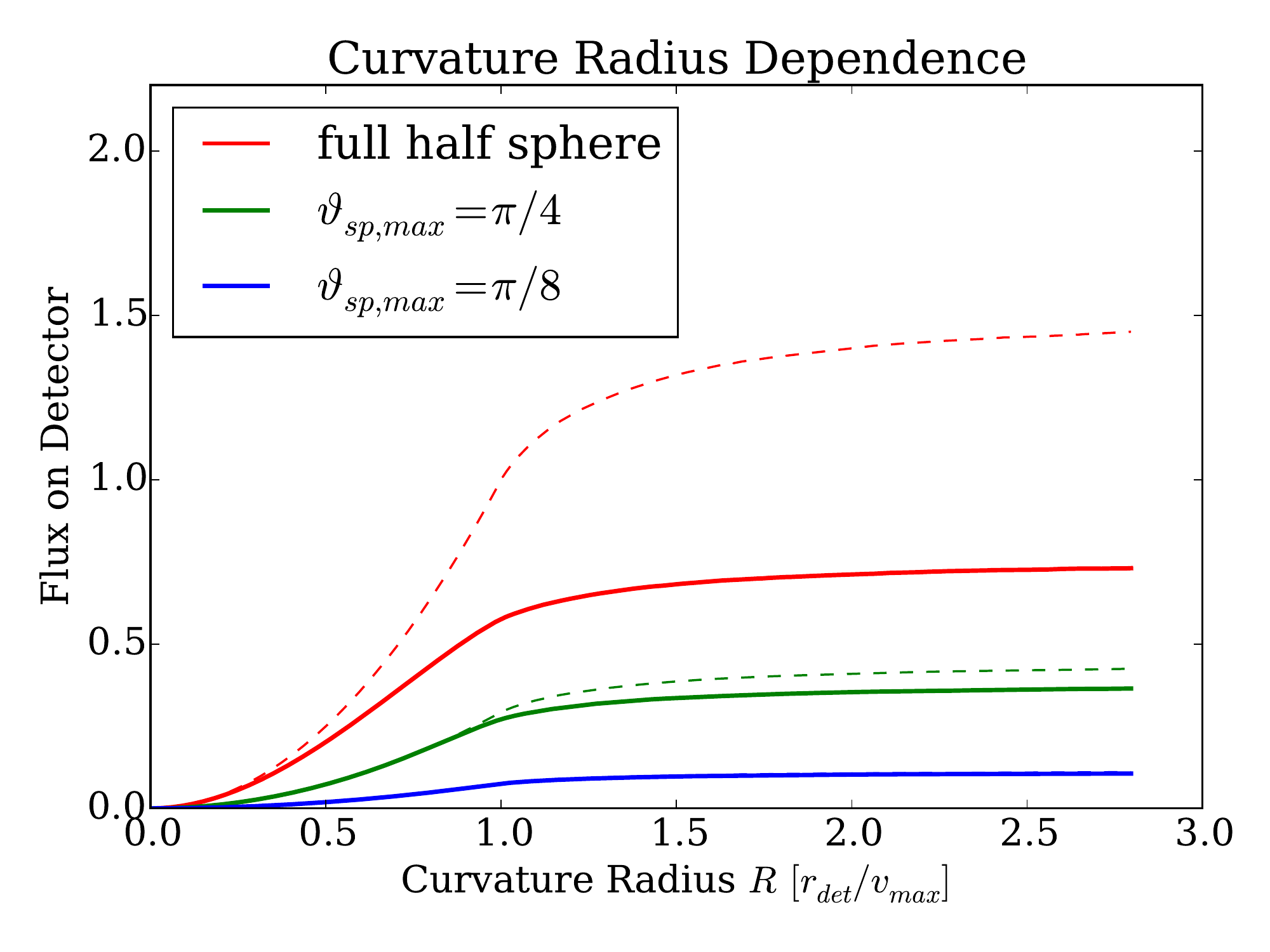} 
 \hspace*{0.5cm}
   \includegraphics[width=7cm]{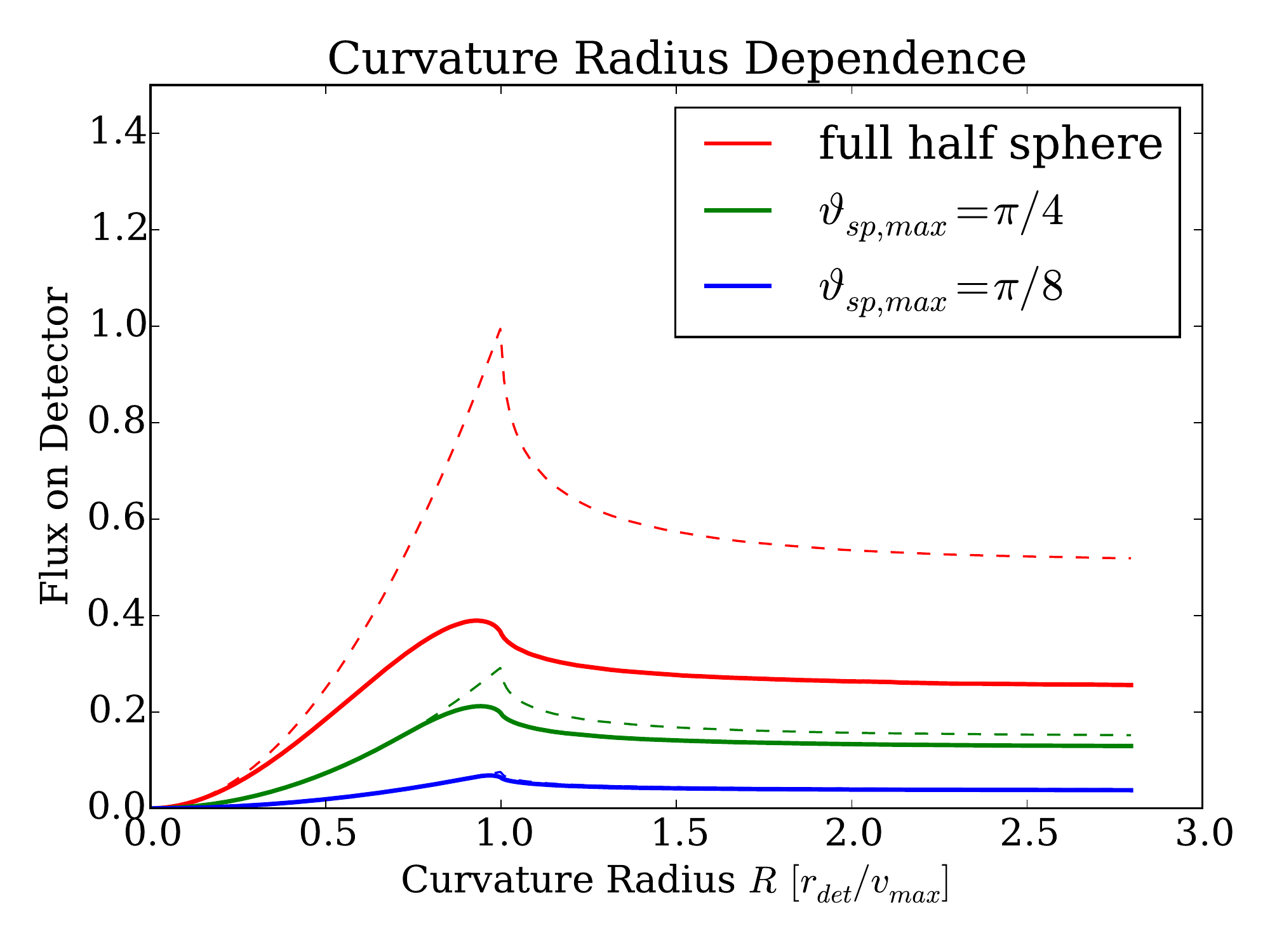} 
   \caption{Total flux of photons impacting a detector of radius $r_{det}$ as a function of the curvature radius for three different opening angles $\vartheta_{sp, max}=\pi/8$, $\vartheta_{sp, max}=\pi/4$, $\vartheta_{sp, max}=\pi/2$. Dashed lines show the results assuming ideal imaging.
   The flux is normalized to the total number of photons produced by a full half sphere of radius $r_{det}/v_{max}$.
   {\bf Left Panel:} Dark matter distribution as given in Eq.~\eqref{uniform}. {\bf Right Panel:} All dark matter particles have velocity $|\mathbf{v}|=v_{max}$ but all directions are equally probable.}
   \label{optimize}
\end{figure}

To increase the photon flux impacting a detector of a given size $r_{det}$ it makes sense to optimize the curvature radius.
This is shown in Fig.~\ref{optimize}. In the left panel we show the result for a dark matter distribution as in Eq.~\eqref{uniform}, whereas in the right panel
all dark matter particles are assumed to have the same velocity $|\mathbf{v}|=v_{max}$ but all directions being equally probable.
Comparing the two we see that increasing the size of the conversion surface is not always beneficial. This is because a larger radius of
curvature leads to larger shifts of the photons away from the centre. Therefore, for a fixed detector size $r_{det}$, only photons with $|\mathbf{v}_{\parallel}|\lesssim r_{det}/R$ will register on the detector, photons with larger velocities will be lost. For a dark matter distribution where most particles have velocities smaller than $v_{max}$ we think that a curvature radius of $R\sim r_{det}/v_{max}$ seems a reasonable choice. Beyond this value
the gain is limited and the increase could even be detrimental.
  
\section{Improved directional resolution with a parabolic mirror}\label{parabolic}
\begin{figure}[t]
\centering
   \includegraphics[width=10cm]{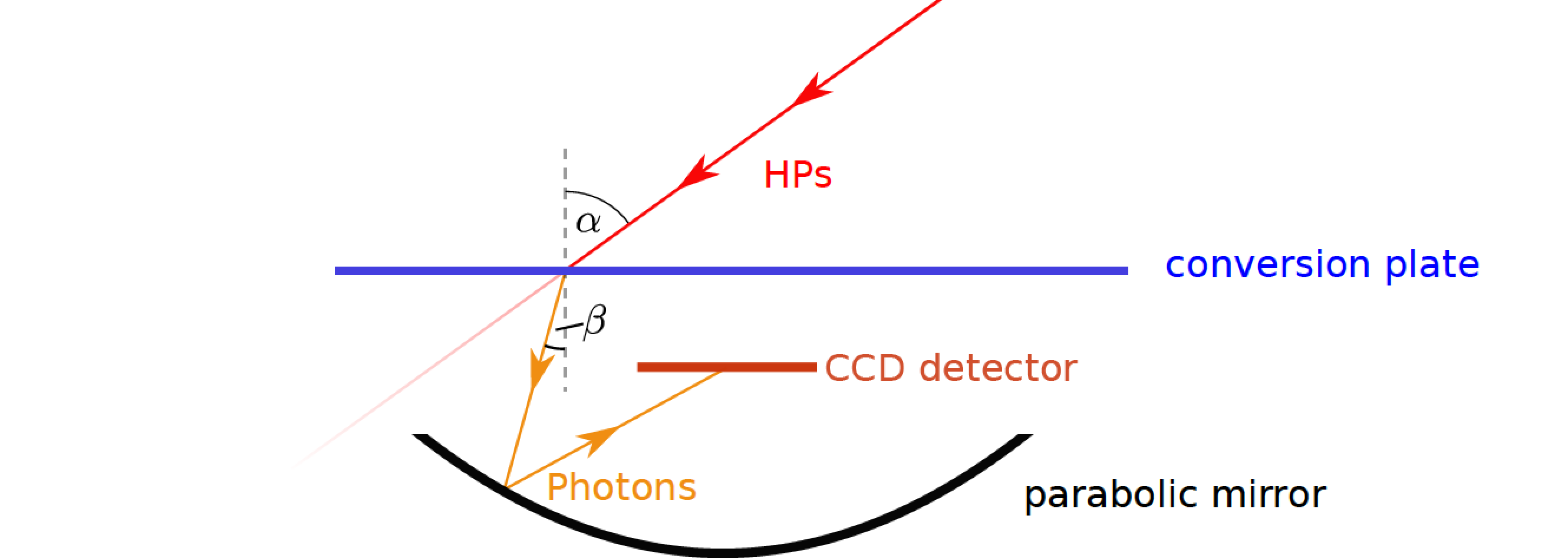} 
   \caption{Sketch of a setup consisting of a reflective plane for the conversion of dark matter WISPs and a parabolic mirror used
   for focussing the produced photons on the detector.}
   \label{parplane}
\end{figure}

Recently a new setup that uses a parabolic mirror and a plane instead of a single spherical mirror has been suggested~\cite{suzuki,redondo}\footnote{The use of a dielectric mirror instead of a simple reflective plane may improve sensitivity~\cite{Jaeckel:2013eha,redondo}.}. The setup is sketched in Fig.~\ref{parplane}. Similarly one could also imagine a setup using a reflecting plane and a focussing lens~\cite{redondo}.

This setup has been devised to employ widely available parabolic mirrors. However, it also provides improved imaging
properties for directional detection.

The main advantage is that the photons emitted from the plane form a plane wave at an angle that is solely determined by the velocity
parallel to the plane. At order $\sim v$ the velocity perpendicular to the plane plays no role. Indeed the first correction only appears at order $v^3$ as can be seen from Eq.~\eqref{inout},
\begin{equation}
\sin(\beta)=\frac{v}{\sqrt{1+v^2}}\sin(\alpha)=v\sin(\alpha)-\frac{v^3}{2}\sin(\alpha)+\ldots=|\mathbf{v}_{\parallel}|-\frac{v^2}{2}|\mathbf{v}_{\parallel}|+\ldots,
\end{equation}
where $\alpha$ is the angle of the incident dark matter particles and $\beta$ the angle of the outgoing wave, as shown in Fig.~\ref{geometry}.

The problem of measuring the velocity of the dark matter parallel to the plane, $\mathbf{v}_{\parallel}$ therefore reduces to that of imaging very distant objects. This is exactly what a telescope does. 
Therefore one can apply standard telescope technology which is very well developed and therefore provides a huge advantage.
In practice we need a telescope with an angular field of view determined by the maximal velocity of the dark matter particles $\sim v_{max}\sim 10^{-3}$.

The simplest suitable telescope is simply a parabolic mirror with a detector in the focus point\footnote{One may wonder if the conversion photons emitted by the mirror itself will cause problems, but this is not the case since the conversion photons emitted from the parabolic mirror itself will be concentrated at roughly two times the focal length.}. 
Of course this also has non-ideal imaging properties. While rays parallel to the axis are perfectly focussed on a point, those rays coming in at an angle suffer from an aberration called coma.
Without going into details let us note that for not too large $D$ the coma has a size (see, e.g.~\cite{telescope})
\begin{equation}
C=\frac{3\beta D^2}{16 f^2},
\end{equation}
where $\beta$ is the angle with respect to the mirror axis, $f$ is the focal length and $D$ is the diameter of the parabolic mirror. 
Taking the size of the coma to be the uncertainty in the determination of the velocity we have
\begin{equation}
\frac{|\Delta\mathbf{v}_{\parallel}|}{|\mathbf{v}_{\parallel}|}=\frac{3D^2}{16 f^2}.
\end{equation}
Let us take a parabolic mirror with an area and focal length of the same size as the FUNK mirror, $f=R/2=1.7\,{\rm m}$ and $A=13\,{\rm m}^2$.
For this setup we find
\begin{equation}
\frac{|\Delta\mathbf{v}_{\parallel}|}{|\mathbf{v}_{\parallel}|}\sim 0.25.
\end{equation}
Already at first glance this looks like an improvement compared to the spherical setup.

Beyond the numerical advantage by a factor of $\sim 2$ when naively comparing to Eq.~\eqref{uncert} there is a perhaps more important advantage.
The uncertainty in the parallel velocity only depends on the parallel velocity itself, the perpendicular velocity of the dark matter particles plays no role. 
It is a purely geometric effect of the imaging system.
In contrast, the smearing in Eq.~\eqref{uncert} is determined by the perpendicular velocity of the incoming dark matter particles.
The purely geometric aberration in the plane-parabolic setup is much easier to correct (and techniques can be directly borrowed from astronomy), as it does not depend on the unknown perpendicular velocity of the dark matter particles.

\section{Conclusions}\label{conclusions}
Reflecting surfaces can convert WISPy dark matter particles such as hidden photons or axion-like particles into photons which can then be detected.
This technique has discovery potential and first experiments are already underway~\cite{Suzuki:2015sza,Dobrich:2014kda}.
The emission angle of the photon depends on the velocity of the dark matter particles, opening opportunities for directional detection. Yet, this also provides a challenge for discovery experiments since the different emission angles will spread out the photons, requiring a sufficiently large detector area. In this note we have investigated how the spread is affected by the size of the employed mirrors.
For spherical mirrors whose size corresponds to sizeable opening angles $\tan(\vartheta_{sp,max})\sim 1$, non-ideal imaging effects can not be neglected for discovery experiments in the calculation of the sensitivity and the velocity resolution for the dark matter particles is limited.
Directional resolution could of course be improved by reducing the mirror opening angles, but this also leads to a loss of signal strength per detector area. One potential alternative is a setup consisting of a reflecting plane, that is used for the conversion of the dark matter particles, and some focussing optics similar to a telescope. Already the simplest setup with a plane and a parabolic mirror has a reduced photon spread and better directional resolution. Further improvement based on standard techniques to correct for aberrations in telescopes seem very possible.

\subsubsection*{Acknowledgements}
JJ would like to thank J.~Redondo for very useful discussions.  JJ gratefully acknowledges support by the Transregio TR33 ``The Dark Universe''.


\begin{thebibliography}{99}

\bibitem{Bertone:2004pz}
  G.~Bertone, D.~Hooper and J.~Silk,
  Phys.\ Rept.\  {\bf 405} (2005) 279
  [hep-ph/0404175].



\bibitem{Arias:2012az}
  P.~Arias, D.~Cadamuro, M.~Goodsell, J.~Jaeckel, J.~Redondo and A.~Ringwald,
  JCAP {\bf 1206} (2012) 013
  [arXiv:1201.5902 [hep-ph]].

\bibitem{Jaeckel:2010ni}
  J.~Jaeckel and A.~Ringwald,
  Ann.\ Rev.\ Nucl.\ Part.\ Sci.\  {\bf 60} (2010) 405
  [arXiv:1002.0329 [hep-ph]].



\bibitem{Horns:2012jf}
  D.~Horns, J.~Jaeckel, A.~Lindner, A.~Lobanov, J.~Redondo and A.~Ringwald,
  JCAP {\bf 1304} (2013) 016
  [arXiv:1212.2970 [hep-ph]].



\bibitem{Suzuki:2015sza}
  J.~Suzuki, T.~Horie, Y.~Inoue and M.~Minowa,
  JCAP {\bf 1509} (2015) 09,  042
  [arXiv:1504.00118 [hep-ex]].


\bibitem{Dobrich:2014kda}
  B.~D\"obrich, K.~Daumiller, R.~Engel, M.~Kowalski, A.~Lindner, J.~Redondo and M.~Roth,
  arXiv:1410.0200 [physics.ins-det].



\bibitem{Jaeckel:2013sqa}
  J.~Jaeckel and J.~Redondo,
  JCAP {\bf 1311} (2013) 016
  [arXiv:1307.7181 [hep-ph]].



\bibitem{Knirck}
S.~Knirck, Bachelor thesis, Heidelberg 2014.

\bibitem{Nelson:2011sf}
  A.~E.~Nelson and J.~Scholtz,
  Phys.\ Rev.\ D {\bf 84} (2011) 103501
  [arXiv:1105.2812 [hep-ph]].



\bibitem{Holdom:1985ag}
  B.~Holdom,
  Phys.\ Lett.\ B {\bf 166} (1986) 196.

\bibitem{Jaeckel:2013ija}
  J.~Jaeckel,
  Frascati Phys.\ Ser.\  {\bf 56} (2012) 172
  [arXiv:1303.1821 [hep-ph]].


\bibitem{Dienes:1996zr}
  K.~R.~Dienes, C.~F.~Kolda and J.~March-Russell,
  Nucl.\ Phys.\ B {\bf 492} (1997) 104
  [hep-ph/9610479];
  S.~A.~Abel, J.~Jaeckel, V.~V.~Khoze and A.~Ringwald,
  Phys.\ Lett.\ B {\bf 666} (2008) 66
  [hep-ph/0608248];
  S.~A.~Abel, M.~D.~Goodsell, J.~Jaeckel, V.~V.~Khoze and A.~Ringwald,
  JHEP {\bf 0807} (2008) 124
  [arXiv:0803.1449 [hep-ph]];
  M.~Goodsell, J.~Jaeckel, J.~Redondo and A.~Ringwald,
  JHEP {\bf 0911} (2009) 027
  [arXiv:0909.0515 [hep-ph]];
  M.~Cicoli, M.~Goodsell, J.~Jaeckel and A.~Ringwald,
  JHEP {\bf 1107} (2011) 114
  [arXiv:1103.3705 [hep-th]];
M.~Goodsell, S.~Ramos-Sanchez and A.~Ringwald, 
  JHEP {\bf 1201} (2012) 021
  [arXiv:1110.6901 [hep-th]].


\bibitem{suzuki}
  J.~Suzuki, Y.~Inoue, T.~Horie and M.~Minowa,
  arXiv:1509.00785 [hep-ex].

\bibitem{redondo}
J.~Redondo, private communication.

\bibitem{Jaeckel:2013eha}
  J.~Jaeckel and J.~Redondo,
  Phys.\ Rev.\ D {\bf 88} (2013) 11,  115002
  [arXiv:1308.1103 [hep-ph]].

\bibitem{telescope}
V.~Sacek, http://www.telescope-optics.net/coma.htm
  
 \end{thebibliography}
\end{document}